\documentclass[prl,twocolumn,graphicx,amssymb,floatfix]{revtex4}
\usepackage{graphicx}
\begin{document}

\title{How to Measure Magnetic Flux with a Single Position Measurement?}
\author{E. Cohen$^1$, L. Vaidman$^1$,  Y. Aharonov$^{1,2}$}
\affiliation{ $^1$Raymond and Beverly Sackler School of Physics and Astronomy\\
 Tel-Aviv University, Tel-Aviv 6997801, Israel\\
 $^2$Schmid College of Science and Technology\\
 Chapman University, Orange, CA 92866, USA}

\begin{abstract}
Current methods for measuring magnetic flux are based on performing many measurements over a large ensemble of electrons. We propose a novel method based on wavefunction ``revival'' for measuring the flux modulo $\frac{hc}{2e}$ using only a single electron. A preliminary analysis of the feasibility of the experiment is provided.

\end{abstract}
\maketitle

In classical theory, physical variables can, in principle, be found by a measurement which consists of just a single event. For example, to measure a field in a particular place we can send a particle in a known initial state (position and velocity) and then deduce the value of the field from the place the particle lands on a screen.  Uncertainty relations  of quantum mechanics put some constrains, however, by increasing the coupling strength we can reach an arbitrary precision also when we use a quantum probe, i.e. a quantum particle. This is a consequence of the correspondence principle.

Some quantum experiments do not have a corresponding classical counterpart. For example, the magnetic Aharonov-Bohm (AB) effect \cite{AB} allows to measure a flux of magnetic field (modulo a constant) through a ring without actually measuring the magnetic field inside the ring. In fact, this experiment has been performed in many systems, from normal metals \cite{resist1,resist2} to 2D electron gas wafer \cite{Heib}, topological insulator nano-wires \cite{Bardarson} and quantum rings in graphene \cite{Russo,Nguyen}.

\begin{figure}[b]
 \begin{center} \includegraphics[width=8.5cm]{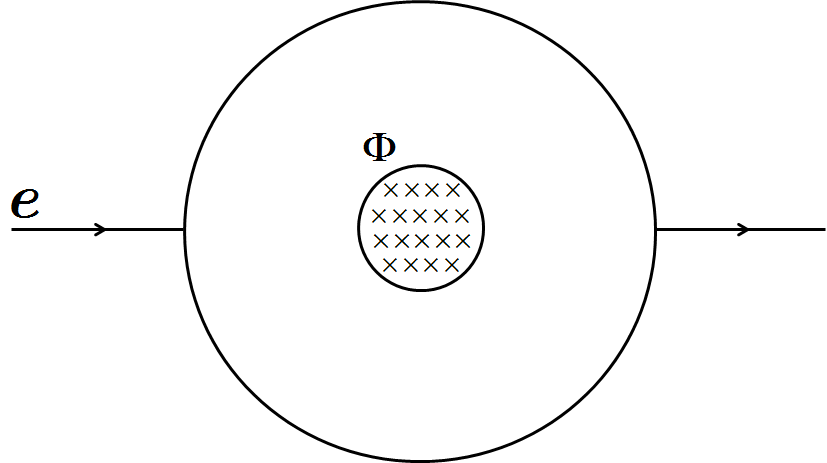}\\
    \caption{Illustration of the typical setup for measuring the AB effect. A coherent beam of electrons enters from the left, splits around the magnetic flux $\Phi$ and recombines. The intensity of the current on the right reveals the modular flux \cite{resist3}. } \end{center}
\end{figure}

In a ring, small enough so that the electron states are not randomized by inelastic scattering during the traversal of the arms of the ring, electronic interference influences the resistance, and thus oscillations of the magnetoresistance as a function of the flux through the ring can be found \cite{resist2}. After calibration, measurement of resistance tells us the value of the flux modulo $\frac{hc}{2e}$ (see Fig. 1). Any measurement of resistance involves measurement of electric current and thus involves measurement of a macroscopic number of electrons. Quantum mechanics allows a direct measurement of a magnetic flux (without measurement of the magnetic field everywhere), but current methods require measurement on an ensemble of electrons encircling the ring with a flux.

\begin{figure}[b]
 \begin{center} \includegraphics[width=6cm]{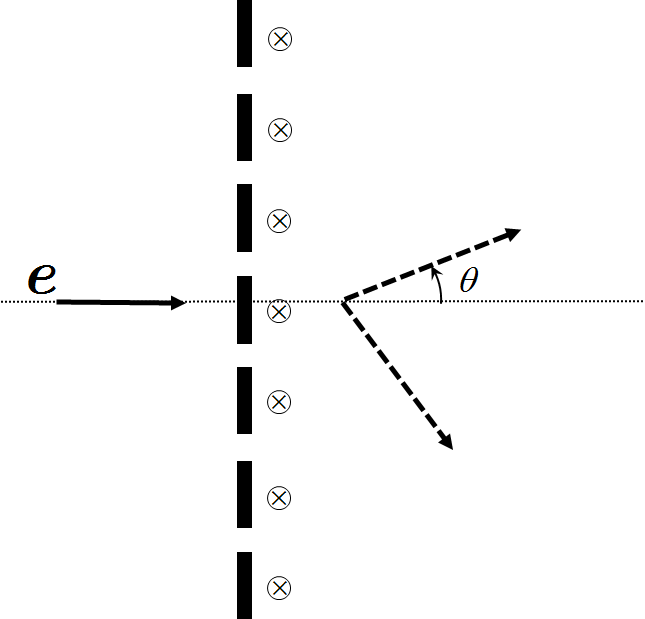}\\
    \caption{Interference of a single electron on a grating with identical solenoids. The angle of the outcoming electron gives the modular flux.} \end{center}
    \end{figure}

Another way to use the AB effect for direct measurement of the flux is to observe the shift in the two-slit interference pattern if we introduce the flux between the two slits. But again, we need an ensemble of measurements to observe the shift of the interference pattern, and a {\it single} electron cannot tell us much about the value of the shift. The shift becomes observable in a {\it single} electron measurement if instead of two slit interference experiment we perform  multiple slit experiment with identical fluxes between every two slits (see Fig. 2) \cite{Aharonov}.
If the slits are spaced a distance $d=\lambda$ apart, where $\lambda$ is the wavelength of the electron, the
interference pattern will consist of two well localized lines at angles corresponding to $\sin \theta_1 =(\Phi/\Phi_0)~ \text {mod}~1$ and $\sin \theta_2 =(\Phi/\Phi_0)~\text{mod}~1~-1$, where $\Phi_0=\frac{hc}{2e}$. Detection of the electron in any of the lines will reveal $\Phi~\text{mod}~\Phi_0$.

The question we want to answer in this letter is how to measure the flux of a {\it single} solenoid, performing a {\it single} experiment with a {\it single} electron.

In our experiment the electron is free to move inside a ring of radius $R$. The width is small so that we can consider the radius of the electron motion to be well defined. We want to measure the magnetic flux through the ring. Let us first consider the case when there is no flux.

We start with the electron well localized in a particular place in the ring. Immediately after its preparation, the wavefunction of the electron will spread out everywhere along the ring. But after a particular time there will be a revival: it will be localized again in the original place.
Indeed, within the ring, the angular part of the Hamiltonian is
\begin{equation}
H=\frac{p_{\varphi}^2}{2mR^2},
\end{equation}
and the angular momentum of the electron is quantized with eigenstates $|n\rangle \equiv |p_{\varphi}=n\hbar \rangle$. The energy eigenstates are
\begin{equation}
E_n=\frac{{\hbar}^{2}n^{2}}{2mR^{2}},
\end{equation}
so after time
\begin{equation}\label{T}
T=\frac{4\pi mR^{2}}{\hbar},
\end{equation}
every angular momentum eigenstate will acquire the same phase. Any state which can be decomposed into a superposition
 \begin{equation}\label{eigen_L}
 |\Psi\rangle=\sum_{n}a_n |n\rangle, ~~~~~~~~~~~~~~|n\rangle\equiv\frac{1}{\sqrt {2\pi}}e^{in\varphi},
\end{equation}
thus, after time $T$, every state will return to the original state (up to an overall phase).

It is of interest to consider the state of the electron after time $t=\frac{T}{2}=\frac{2\pi mR^{2}}{\hbar}$. For even $n$ we obtain $|n\rangle \rightarrow |n\rangle $, while for odd $n$ we have $|n\rangle \rightarrow -|n\rangle $. This is exactly what happens if we rotate the state by $\pi$, i.e. $\varphi \rightarrow \varphi+\pi$. Hence, at time $T/2$ the electron should be found at $\varphi=\pi$. Using the same reasoning, it can be shown that after time $t=T/3$ the particle is in a superposition of being localized around $\varphi=0$, $\varphi=2\pi/3$ and $\varphi=4\pi/3$. After time $t=T/4$ it is in a superposition of being localized around $\varphi=0$ and $\varphi=\pi$ and so on for every $T/k$, when the superposition at odd $k$ instances is ``richer'' than the superposition at even $k$ instances). The dynamics hence describes a well localized wave packet at $t=0$ which quickly spreads over the whole ring and acquires specific positions at certain times, until coming back to its starting point at $t=T$.

This kind of dynamics is known as ``revival'' \cite{Eberly}. Recently, ``fractional revival'', like the one demonstrated above, was also suggested \cite{Romera}. However, the analysis we hereby propose is unique in utilizing this phenomenon for gathering information about the setup (magnetic flux in this case) using a single particle.

Let us consider now the AB setup in which there is a flux $\Phi$ inside the ring, see Fig. 3. We can use the Coulomb gauge such that the Hamiltonian is
 \begin{equation}\label{Hamiltonian}
 H=\frac{(p_{\varphi}+\frac{e}{2\pi c}\Phi)^{2}}{2mR^{2}},
\end{equation}
with energies
 \begin{equation}
E_n=\frac{(\hbar n+ \frac{e}{2\pi c}\Phi)^{2}}{2mR^{2}}.
\end{equation}
At $t=T=\frac{4\pi mR^{2}}{\hbar}$ each eigenstate $|n\rangle$ will be multiplied by
 \begin{equation}
e^{-iE_nT/\hbar}=e^{i \cdot const} e^{-2\pi n^2 i} e^{-2n (e\Phi/c\hbar) i},
\end{equation}
where the first exponent is $n$-independent and the second equals $1$. Hence, apart from an overall phase, every eigenstate $|n\rangle$ will be multiplied by $e^{2in\phi_{AB}}$, where $\phi_{AB}=\frac{e}{ c\hbar }\Phi$ is the well known AB phase \cite{AB}. This is equivalent to the application of the shift operator $e^{2ip_\varphi\phi_{AB}/\hbar}$, and therefore, after time $T$, the electron will be localized at angle $\varphi=2\phi_{AB}$, from which the flux modulo $\frac{hc}{2e}$ can be found. This stands in contrast with the usual methods of measuring flux which requires measurements over a large ensemble, and enables to test the AB effect in a completely new manner.

\begin{figure}[b]
 \begin{center} \includegraphics[width=8.5cm]{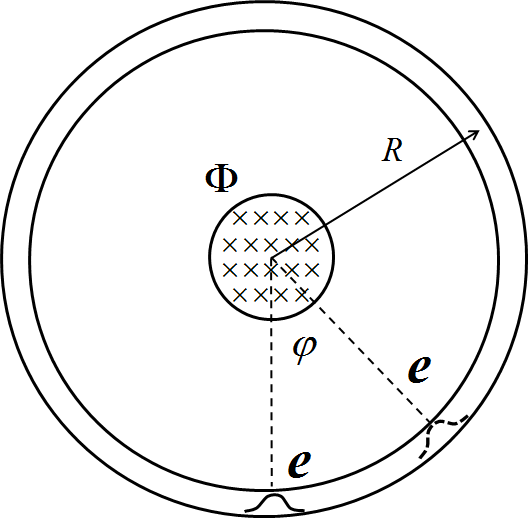}\\
    \caption{Measuring flux with a single electron. The electron, initially localized around $\varphi =0$, ends up at time $T=\frac{4\pi mR^{2}}{\hbar}$ localized at the angle $\varphi=2\phi_{AB}$, from which the modular flux can be extracted. }\end{center} \label{flux}
\end{figure}

But how precise is the measurement? Ideally, we would have wanted to use for the well localized electronic wavepacket a delta function. However, to reach a finite energy cost, we will use a Gaussian truncation with zero mean
\begin{equation}
\Psi(\varphi)=\frac{1}{2\pi}\sum_{n=-\infty}^{\infty}e^{-n^2/(\Delta n)^2}e^{in(\varphi-\varphi_0)},
\end{equation}

The error in the angular position measurement is \cite{Harter}
\begin{equation}
\Delta \varphi\approx\frac{1}{\pi\Delta n},
\end{equation}
and the error in measuring the flux will be $\Delta\Phi=\frac{c\hbar \Delta \varphi}{2e}$. A relative error of $0.5\%$,  with respect to one flux quantum $\frac{hc}{2e}$,  would require a lower bound on the initial uncertainty, $\Delta n \ge 10$.


Let us now take into account relativistic corrections.
The first order correction for the energy eigenvalues is \cite{LL}

\begin{equation}
\delta E_n=\frac{{(\hbar n)}^4}{8c^2m^3R^4}.
\end{equation}
After time $T$, the additional phase shift of the $n$th level equals
\begin{equation}
\delta \varphi_n=\frac{\pi{\hbar}^2 n^4}{2(cmR)^2}.
\end{equation}
Requiring $\delta \varphi_n<<\Delta \varphi$, for  $|n| \le 10$, we find a relativistic limitation on the radius
\begin{equation}
R >> \frac{\pi \hbar}{mc}\sqrt{\frac{(\Delta n)^5}{2}}=3\cdot 10^{-10} {\text m}.
\end{equation}

For large $|n|$, another important factor is the centrifugal force which changes the electron radial wavefunction. The effective radius of motion might be different for different $n$ and it will cause a shift in phase. For estimating this effect one has to know the bounding potential.
If, for example, the radial potential is linear, the shift in the radius of motion can be analytically measured as was performed in \cite{Olendski}, but it does not seem like a realistic model for actual experiments. In realistic cases, the binding potential becomes very steep towards the edge, thereby providing approximately a fixed radius.




In recent experiments where the AB phase was measured \cite{Heib,Khatua}, the radius of motion was $R \simeq 10^{-6} {\text m}$ which corresponds to $T \approx 10^{-7} {\text s}$. It seems feasible to perform an experiment with this time resolution. The reason for using a small radius in a laboratory experiment, is that the main limiting factor is the coherence length of the electrons moving in the ring. To reach high coherence length the temperature of the electron gas has to be very low (140 mK was used in \cite{Heib}).



An idea for a possible physical implementation, is to use edge states which have been recently employed to demonstrate the AB effect in quasi 1D nanoribbons \cite{Peng} and in 2D topological insulators \cite{Gusev}. We hope that we can see the revival effect in these edge states. Suitable gates can prepare, and later find, the initial and final states. To prepare the electron in a localized state, we need to account for the chirality of edge states. Therefore, we have to prepare a wavepacket with Gaussian truncation of large, positive mean so that only positive momenta will contribute. The mean rotation effect can be eliminated by arranging that the electron will complete
an integer number of lapses during time $T$.



We have proposed in this letter a method which allows indirect measurement of flux using detection of a single electron experiencing the AB effect. This stands in contrast with all other experiments performed to date, where magnetic flux measurements were based on the AB effect of large ensembles. Our method utilizes the fact that due to the special form of the energy spectrum the wavefunction of the electron undergoes a revival, and thus returns to be well localized. Preliminary considerations suggest the feasibility of an experimental realization of our proposal.


{\bf Acknowledgements.}
We are thankful to Victor Fleurov, Moti Heiblum, Yuval Ronen and Eran Sela for helpful comments and discussions. This work has been supported by the Israel Science Foundation Grant No. 1311/14, by the ICORE Excellence Center ``Circle of Light'' and by DIP, the German-Israeli Project cooperation.



\end{document}